\documentclass[twocolumn,amsmath,trackchanges]{aastex702}

\begin{document}

\title{Inverting and Up-channelizing Critically-Sampled Polyphase Filter Banks}

\author[orcid=0009-0000-5154-3425]{Stephen Fay}
\affiliation{Department of Physics, McGill University, Montreal, Quebec H3A 2T8, Canada}
\email{stephen.fay@mail.mcgill.ca}  

\author[orcid=0009-0002-1555-1668]{Mohan Agrawal}
\affiliation{Department of Physics, McGill University, Montreal, Quebec H3A 2T8, Canada}
\affiliation{Trottier Space Institute, McGill University, Montreal, Quebec H3A 2A7, Canada}
\email{mohan.agrawal@mail.mcgill.ca}   

\correspondingauthor{Mohan Agrawal}
\email{mohan.agrawal@mail.mcgill.ca}   

\author[orcid=0000-0002-4098-9533]{H.~Cynthia Chiang}
\affiliation{Department of Physics, McGill University, Montreal, Quebec H3A 2T8, Canada}
\affiliation{Trottier Space Institute, McGill University, Montreal, Quebec H3A 2A7, Canada}
\email{cynthia.chiang@mcgill.ca}

\author[orcid=0000-0003-1130-6390]{Aman Chokshi}
\affiliation{Department of Physics, McGill University, Montreal, Quebec H3A 2T8, Canada}
\affiliation{Trottier Space Institute, McGill University, Montreal, Quebec H3A 2A7, Canada}
\email{aman.chokshi@mcgill.ca}

\author[orcid=0000-0001-6903-5074]{Jonathan Sievers}
\affiliation{Department of Physics, McGill University, Montreal, Quebec H3A 2T8, Canada}
\affiliation{Trottier Space Institute, McGill University, Montreal, Quebec H3A 2A7, Canada}
\email{jonathan.sieversi@mcgill.ca}  

\author[orcid=0009-0006-8752-1424]{Simon Tartakovsky}
\affiliation{Department of Physics, Princeton University, Princeton, New Jersey 08540, USA}
\email{st5640@princeton.edu} 

% \collaboration{all}{The Terra Mater collaboration}

%% Use the \collaboration command to identify collaborations. This command
%% takes an optional argument that is either a number or the word "all"
%% which tells the compiler how many of the authors above the command to
%% show. For example "\collaboration[all]{(DELVE Collaboration)}" wil include
%% all the authors above this command.
%%
%% Mark off the abstract in the ``abstract'' environment. 
\begin{abstract}

Polyphase filter banks (PFBs) are widely used to channelize digitized time-domain data in real time. However, applications such as radio astronomy often require higher spectral resolution than real-time systems can provide. We introduce a circulant inverse-PFB formalism and present a new, scalable, fast Fourier transform (FFT)-based algorithm for inverting critically sampled PFBs. This formalism explicitly identifies the origin of quantization noise amplification and enables two complementary mitigation strategies: (1) a fast, practical Wiener filter and (2) a rigorous maximum-likelihood reconstruction with time-domain priors. We evaluate both approaches using simulated 4-bit quantized PFB data. For autospectra, Wiener filtering confines reconstruction errors to less than 10\% of the channel width, with a peak error below 10\%. The maximum-likelihood approach further reduces reconstruction errors as additional time-domain prior information is incorporated, with priors spanning 10\% of the time-domain samples reducing worst-case errors to below 2\%. This framework enables high-resolution spectral analysis of archival PFB data, facilitating applications including long-baseline interferometry, pulsar and fast radio burst searches, and ultra-narrow-band physics experiments.

\end{abstract}

%% Keywords should appear after the \end{abstract} command. 
%% The AAS Journals now uses Unified Astronomy Thesaurus (UAT) concepts:
%% https://astrothesaurus.org
%% You will be asked to selected these concepts during the submission process
%% but this old "keyword" functionality is maintained in case authors want
%% to include these concepts in their preprints.
%%
%% You can use the \uat command to link your UAT concepts back its source.
% \keywords{\uat{Galaxies}{573} --- \uat{Cosmology}{343} --- \uat{High Energy astrophysics}{739} --- \uat{Interstellar medium}{847} --- \uat{Stellar astronomy}{1583} --- \uat{Solar physics}{1476}}

%% From the front matter, we move on to the body of the paper.
%% Sections are demarcated by \section and \subsection, respectively.
%% Observe the use of the LaTeX \label
%% command after the \subsection to give a symbolic KEY to the
%% subsection for cross-referencing in a \ref command.
%% You can use LaTeX's \ref and \label commands to keep track of
%% cross-references to sections, equations, tables, and figures.
%% That way, if you change the order of any elements, LaTeX will
%% automatically renumber them.

\section{Introduction}\label{sec: intro}

A polyphase filter bank \citep[PFB;][]{harris2003digital,crochiere2005interpolation,vaidyanathan2002multirate,bellanger1976digital,price2021spectrometers} is a signal processing technique, common in fields like radio astronomy and telecommunications, for splitting a broadband digital signal into many narrow frequency channels. PFB channelizers allow custom sculpting of the frequency response of the output channels with user-defined filters while also being much faster than traditional channelizing schemes. Their distinct advantage is manifest in radio astronomy where minimizing spectral leakage \citep{oppenheim1999discrete} between adjacent channels is necessary to prevent strong human-generated interference from contaminating frequencies of scientific interest. A PFB channelizer can provide any desired level of passband size and sidelobe suppression with an appropriate length of filter used in its construction. Typical PFBs used in modern radio telescopes provide excellent stop-band attenuation (Fig.~\ref{fig: sidelobes}), and for the same level attenuation, do so at a fraction of the computational cost associated with naive channelization techniques.

Radio telescopes typically process large instantaneous radio-frequency (RF) bandwidths, often ranging up to several GHz. Processing large RF bandwidths requires fast sampling rates, which results in high throughput data rates. For this reason, hardware constraints at the input, such as buffer size limitations on field programmable gate arrays (FPGA), or at the output, such as data transmission or storage capacity, limit the frequency resolution achievable with real-time channelization. Therefore, many instruments record only coarsely-channelized PFB data, even though finer frequency resolution could be scientifically valuable. 

Recovering the original timestream from archival PFB data lifts these restrictions: once inverted, the data can be re-channelized to arbitrarily fine frequency resolution to suit the science case at hand. This capability is desirable across a range of applications. In long baseline interferometry, for example, coarse channelization limits achievable frequency resolution, which in turn limits the maximum usable baseline length and field-of-view because of the bandwidth smearing effect \citep{bridle1999bandwidth}. Re-channelization of coarse PFB data mitigates this effect, thereby enabling ultra-wide-field long-baseline interferometry \citep{chiang2020arraylongbaselineantennas}. An accurate PFB inversion scheme could also enhance the sensitivity of pipelines designed for pulsar searches, and fast radio burst search and localization \citep{cho2020spectropolarimetric,leung2025vlbi}. Lastly, re-channelization capability could potentially enable searches for new physics, such as ultra-narrowband (few-Hz) signals from axion-like particles in the ionosphere \citep{beadle2024resonant}, with existing radio telescopes. 

The most common implementation of a PFB channelizer is \textit{critically sampled}, meaning that the frequency spacing between channels is the same as a channel's Nyquist bandwidth. Passing a real-valued signal of bandwidth $B$, sampled at the Nyquist rate of $2B$, through a PFB that uses an $N$-point FFT at the channelization stage produces $n_{chan}=N/2$ positive frequency channels that are spaced $B/n_{chan}$ apart. If the PFB is critically-sampled, then for each channel, a new complex-valued sample also arrives at the rate of $B/n_{chan}$, which makes the input rate and the output rate (summed across all channels) equal. Critical-sampling causes any signal leaking outside a channel's Nyquist width of $B/{n_{chan}}$ to alias back into the channel, corrupting the channel edges (see Fig.~\ref{fig: sidelobes}). This aliasing makes critically-sampled PFBs difficult to invert. As will be discussed later, aliasing alone is not sufficient to make inversion impossible: while information has been scrambled, it is not lost.

Traditionally, invertible PFBs have been designed from the ground-up using so-called \textit{perfect-reconstruction} filters. These filters exactly cancel aliasing artifacts from neighbouring channels, but this property imposes strict constraints on their length and shape \citep{malvar1990modulated, lin1995linear, vaidyanathan2002multirate, vetterli2002perfect}. Such restrictions limit the filter's versatility in providing the desired level of stop-band attenuation and pass-band flatness with the available hardware resources. Certain classes of these filters also lack a linear phase response \citep{smith1984procedure}, which makes them further unsuitable for phase-sensitive applications where linearity is required.

Another way to sidestep the difficulties posed by critical sampling is to use oversampled PFBs \citep[OS-PFBs;][]{morrison2020performance, arnaldi2021oversampled, tuthill2012development}, where the output timestream is sampled faster than $B/n_{chan}$. A higher output sampling rate makes a channel's Nyquist bandwidth larger than the channel width. The out-of-channel ripples thus fall within a channel's Nyquist band, reducing the aliasing. However, the benefits of OS-PFBs come at the cost of additional hardware resources that are required to process the now higher output data rate, and correct for systematic phase rotations introduced by the oversampling \citep{tuthill2015compensating}. 

In this paper, we focus on the challenges associated with inverting critically-sampled PFBs. First we pinpoint the reasons for invertibility issues that fundamentally arise from symmetry properties of the filter weights (windows) typically used in PFBs. We then present a computationally efficient inversion algorithm. The organization of the rest of the paper is as follows. Section~\ref{sec: forward pfb} establishes the foundations of PFBs, from ideal channelizers to their interpretation as independent correlations. Section~\ref{sec: inverse pfb} presents an FFT-based algorithm for fast PFB inversion and addresses the effects of quantization noise on the inverted PFB. Finally, section~\ref{sec: mitigating quantization} augments the inversion algorithm with two de-noising methods: a Wiener filter approach, and a time-domain maximum-likelihood approach, and compares their performance using simulated data.

\section{Polyphase Filter Banks}\label{sec: forward pfb}

\subsection{Ideal Channelizer}

The process of channelizing a timestream signal refers to splitting the information content into narrow frequency bands, or ``channels.'' A band-limited, real timestream $x(t)$ can be split into $n_{chan}$ channels, each of which has its own corresponding timestream.  The sum of these per-channel timestreams yields the original $x(t)$. Ideally, any individual channel $k$ having width $\Delta f$ has information that is strictly contained within the frequency range $|f - f_k| \le \Delta f/2$, with no information content from other frequencies. The ideal channelizer is straightforward.  To obtain the timestream for channel $k$, the
Fourier transform $X(f)$ of the original timestream is calculated and then multiplied by a boxcar filter, defined as $H_k(f) = 1$ for $|f - f_k| \le \Delta f/2$, and zero otherwise. This boxcar simply selects frequencies within the vicinity of $f_k$ (the $k$'th channel) and rejects frequencies outside this channel. By the convolution theorem, multiplying $X(f)$ with a boxcar centred at $f_k$ is equivalent to convolving the timestream with the shifted boxcar's Fourier transform: a sinc function times a phase $e^{j2\pi f_k t}$. %This top-hat multiplication removes all information content at frequencies outside the channel of interest, while preserving all information within the channel. By convolution theorem, multiplication of frequency spectrum by $H_k(f)$ is equivalent to convolution of the timestream $x(t)$ with the Fourier transform of $H_k(f)$, which is simply the $sinc$ function multiplied by the complex exponential $e^{j 2 \pi f_k t}$.

If the bandwidth of $x(t)$ is $B$, such that $B = n_{chan}\Delta f$, then the timestream can be digitized by sampling at the Nyquist rate of $2B$. The convolution for obtaining the timestream of channel $k$ can then be written as a sum over the samples of $x(t)$:
\begin{equation}
  y_k(t) = \sum_m x[t_m] \underbrace{{\rm sinc}(t_m-t) e^{-i 2 \pi f_k (t_m-t)}}_{\text{Convolution with the filter}},
\label{eq:chan_tod}
\end{equation}
where $t_m = m/2B$ are the timestream samples and $y_k(t)$ are the timestream samples for channel $k$. As the frequency content of the resulting $n_{chan}$ channelized timestreams is limited to $\Delta f = B/n_{chan}$, each channelized timestream can be sampled more sparsely than $x(t)$.  The lowest possible sampling rate is set by the Nyquist limit to be $B/n_{chan} = \Delta f$, which reduces the number of output samples of $y_k(t)$ by a factor of $n_{chan}$ relative to the length of $x(t)$. Now, if the channelizer output for channel $k$ is evaluated at times that are integer multiples for $1/\Delta f$, then \eqref{eq:chan_tod} reduces to an infinitely-long discrete Fourier transform (DFT)\footnote{This formulation simplifies the formal boundary mapping between continuous spectra and discrete DFT bins. The underlying mechanism is that sampling the channelized output at the downsampled rate of $B/n_{\text{chan}}$ frequency-translates each channel to baseband, causing the complex phase factor $e^{-j 2\pi \nu_k t}$ to evaluate to unity at these sampling instants.}. Moreover, as the output of the channelizer is now complex, the total data rate across all channels is twice $n_{chan}$ times $B/n_{nchan}$, equal to $2B$, which is the input data rate of $x(t)$. This property makes the channelizer critically sampled.

\subsection{Introduction to Polyphase Filter Banks}

Channelizing a real timestream with finite length $M$ requires some modifications to the ideal channelizer. 
%to obtain a computationally tractable implementation. 
First, the sinc filter is truncated to the same length $M$ as the timestream. To obtain $n_{chan}$ complex-valued output channels, a minimum of $N=2n_{chan}$ real-valued input timestream points are required, such that $M\ge N$. We refer to both a unit of $N$ time-domain samples and a set of $N/2$ complex channelized values as a \textit{frame} of data. A sequence of $p$ consecutive frames is then defined as a \textit{segment} of $M$ samples. As M increases, the Fourier transform of the truncated sinc progressively approaches the ideal boxcar filter. The ratio $p = M/N$ is known as the number of PFB \textit{taps}. In practice, $p$ is typically determined by available computational memory, with $p=4$ being common in radio astronomy. 

With $M = pN$ set as the number of timestream data points to channelize, \eqref{eq:chan_tod} can be rewritten for one segment of data of size $M$ as
\begin{equation}
  y[k] = \sum_{m=0}^{M-1} x[m] {\rm sinc}[m - M/2] e^{-i 2 \pi m k /N}.
\label{eq:chan_dft}
\end{equation}
This form can be obtained by using the following substitutions: $f_k = k B/n_{nchan}$, $t_m = m/2B$, $N = 2n_{chan}$, along with the fact that $t$ is an integer multiple of $n_{chan}/B$. Moreover, as the summation starts from $m=0$, the sinc is shifted by $M/2$ samples so that it is symmetric across one block of data\footnote{A filter having an even frequency response (top-hat), must necessarily be even in time.}. %This is due to the fact that for each output sample, the ideal filter in \eqref{eq:chan_tod} spans symmetrically across negative and positive times. 

If $M>N$, then \eqref{eq:chan_dft} is equivalent to taking a longer $M$-sized DFT, but with only every $p$'th frequency channel retained:
\begin{equation}
  X[k'=pk] = \sum_{m=0}^{M-1} x[m] {\rm sinc}[m - M/2] \underbrace{e^{-i 2 \pi m pk/M}}_{e^{-i 2 \pi m k/N}}.
\label{eq:chan_dft_osamp}
\end{equation}
The efficiency of this processing step can be improved by noting that the exponential in \eqref{eq:chan_dft} repeats with a period of $N$. Thus, a summation over $M=pN$ has $p$ identical exponentials. This redundancy can be exploited by dividing the sum over $M$ samples into $p$ different sums, using the substitution $m = rN + n$, where $r < p$ and $n < N$. Letting $h[m] \equiv {\rm sinc}[m - M/2]$ for notational brevity, it follows that
\begin{equation}
    y[k] = \sum_{n=0}^{N-1} e^{-i 2 \pi m k /N} \sum_{r=0}^{p-1} x[rN + n] h[rN + M].
    \label{eq:pfb}
\end{equation}
Equation \eqref{eq:pfb} represents channelization with a polyphase filter bank. The sequence of operations involved in producing one PFB spectrum, as illustrated in Fig.~\ref{fig: polyphase-structure-3panel}, are as follows. A timestream segment of length $pN$ is multiplied by a sinc and divided into $p$ sequential frames of length $N$, starting at samples $0$, $N$, ...$(p-1)N$ respectively. Corresponding samples of theses $p$ timestreams are summed, and the resulting $N$ samples are Fourier transformed. This process is repeated for the adjacent timestream segment, starting a frame ($N$ samples) later. Each data point is therefore used $p$ times in the computation. The power of the PFB lies in achieving the filtering fidelity of a Fourier transform $p$ times longer than the transform actually computed. For a given number of frequency channels, increasing the number of taps $p$ enhances both the flatness of the response as well as out-of-band rejection. 

\begin{figure}[tb]
    \centering
    \includegraphics[width=1.0\linewidth]{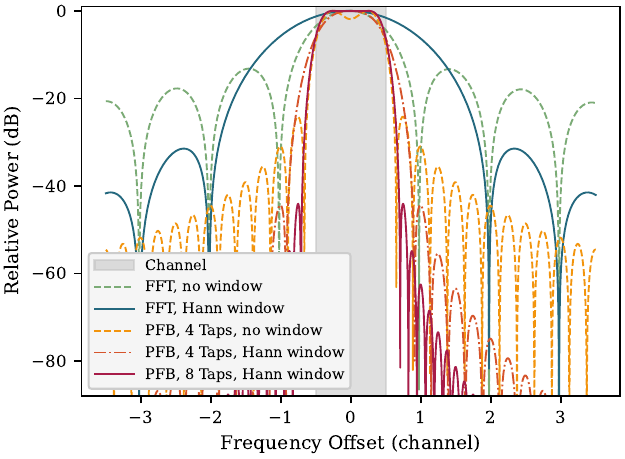}
    \caption{Comparison of frequency responses of a channel of PFB and FFT filter banks. All responses are normalized to unity at the center of the reference channel. The gray shaded area indicates the ideal pass band of a digital channel. The output of a channelizer at any given instant is the input spectrum multiplied by the frequency response. Thus, the plots indicate the amount of power contributed to the reference channel by its neighbors (referred to as spectral leakage). Compared to a simple FFT, a windowed FFT reduces the level of sidelobes at the expense of a wider main-lobe. In contrast, a PFB achieves both a narrow main-lobe and low side-lobes.}
    \label{fig: sidelobes}
\end{figure}

In practice, the filter $h[m]$ is multiplied by a smooth taper (or windowing) function \citep{rabiner1996multirate} to avoid abruptly truncating the sinc, allowing the shapes of the pass and side-bands to be fine tuned. Because the length of the filter is necessarily finite, its response outside the nominal channel width of $B/n_{chan}$ cannot be zero. When the sample rate of a channel timestream is reduced to $B/n_{chan}$, this spectral leakage aliases back into the channel. Fig.~\ref{fig: sidelobes} compares the spectral response of various PFB filters varying in their length and taper functions. In the rest of this text, \textit{window} and \textit{filter} are used interchangeably to mean the sinc function (optionally) multiplied by a conventional taper function. 

\subsection{The PFB as Independent Correlations}

The first step towards PFB inversion is rewriting the channelizer in \eqref{eq:pfb} as a matrix operation:
\begin{equation}\label{eq: pfboperator}
\mathbf{y} = \mathcal{FSW}\mathbf{x}.
\end{equation}
Here, the PFB operator $\mathcal{FSW}$ acts on an input timestream vector, $\mathbf{x}$, of length $pN$. First, the square, diagonal window matrix $\mathcal{W}$ element-wise multiplies $\mathbf{x}$ with the filter weights. Next, $\mathcal{S}$ is a summing matrix of shape $[N, pN]$ that comprises $p$ identity matrices of size $N$ stacked horizontally. This matrix pointwise-sums samples of $p$ consecutive time-domain frames, outputting an $N$-long vector. Finally, $\mathcal{F}$ is a $N$-point DFT operator with $\mathcal{F}_{n,k}=\exp(-j2\pi n k/N)$. To compute the PFB of a longer timestream, the operator $\mathcal{FSW}$ is repeatedly applied to successive timestream segments, each shifted by a frame of $N$ samples. This section provides a detailed breakdown of $\mathcal{FSW}$, which encodes a limited set of independent operations. By reshaping pieces of the constituent matrices, these operations can be isolated from each other, yielding a framework for PFB inversion (and its subsequent implementation with parallel computations).

\begin{figure*}[!htb]
    \centering
    \includegraphics[width=1\linewidth]{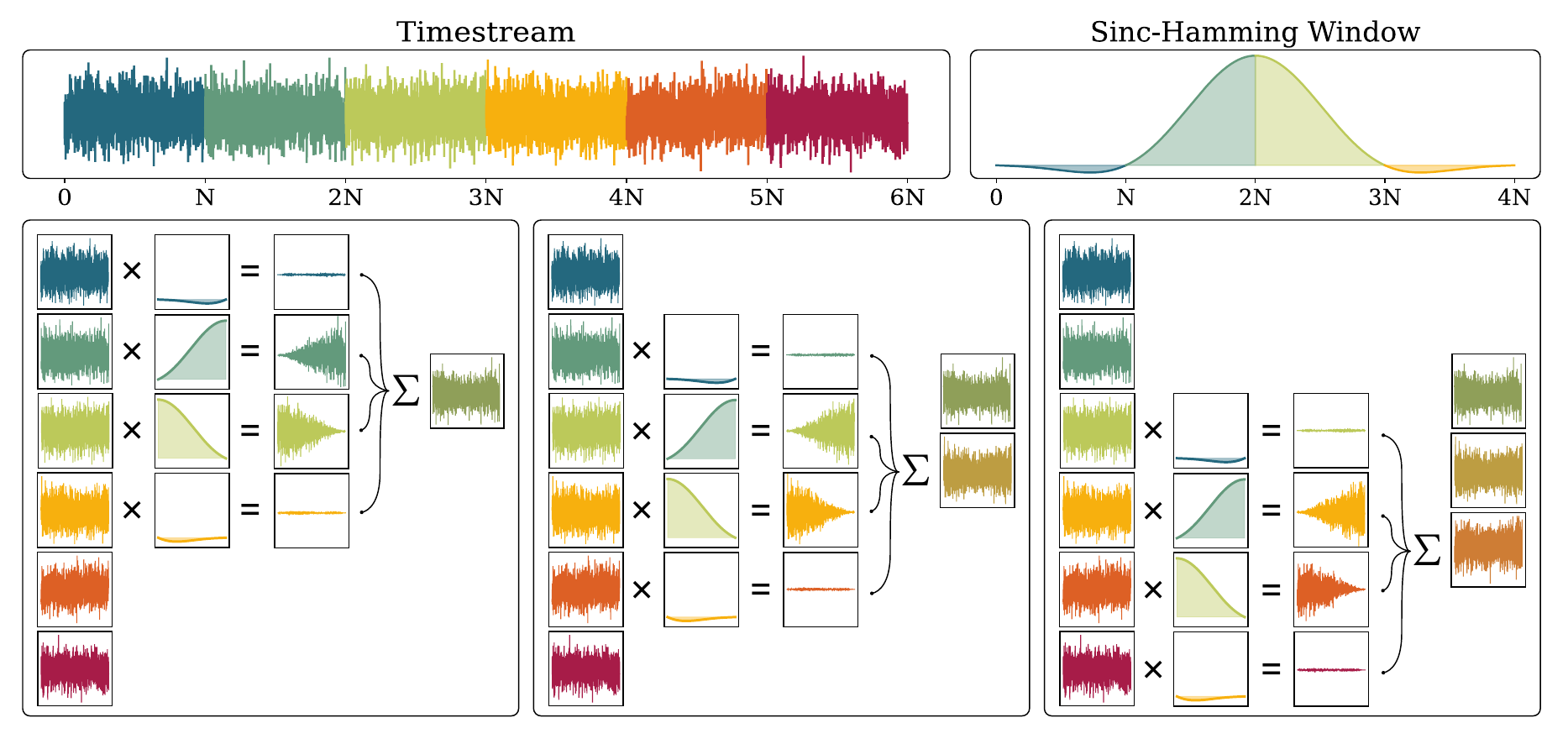}
    \caption{Sequence of steps involved in a PFB, shown without the final Fourier transform. The illustrations demonstrate a 4-tap, $N$-point PFB. The top panels show 6 frames of timestream data, and the $4N$-long filter window composed of sinc function pointwise-multiplied with a Hamming taper function. The three lower panels demonstrate the operations performed by the PFB on three successive overlapping segments of four frames: first, multiply the four frames with a window weights, then sum the transformed frames pointwise in the vertical direction, producing one frame of filtered data. The operation is repeated on an input shifted by a frame. Though not illustrated here, each of the final output frames are Fourier transformed to complete the PFB channelization.}%, (3)  (2) reshaping this segment into an array of shape $(4, N)$, (3) sum vertically along columns, and (4) performing a single, short FFT on the resulting $N$-long segment. We then shift the input segment by $N$ samples and continue channelizing. }
    \label{fig: polyphase-structure-3panel}
\end{figure*}

The only non-zero entries in row $n$ of $\mathcal{SW}$ are: $h[n], h[n+N], \ldots, h[n+(p-1)N]$ in columns $n, n+N, \ldots, n+(p-1)N$, respectively. Thus, each $\mathcal{SW}$ row combines only the corresponding $p$ timestream samples. To streamline the application of $\mathcal{SW}$ to an arbitrarily long timestream, two quantities are reshaped.  First, the input timestream is written as a 2-D array with $N$ columns and an arbitrary number of rows, and second, the $pN$-long window is reshaped into an array of dimensions $[p,N]$.  With this restructuring, a single application of $\mathcal{SW}$ encompasses two steps. First, $p$ rows (frames) of the data array are elementwise multiplied by the window array. Second, all $p$ rows of the resulting product are summed along columns, yielding a $N$-long output which is passed to $\mathcal{F}$, resulting in one PFB spectrum. To obtain the next spectrum, the multiply ($\mathcal{W}$) and add ($\mathcal{S}$) operations are repeated again, but with the data array segment shifted by one frame (row). Repeated application of $\mathcal{SW}$ can be envisioned as the window array sliding down the data array, as illustrated in Fig.~\ref{fig: polyphase-structure-3panel}.

The reshaped representation reduces the $\mathcal{SW}$ operation to independent column-wise multiply-and-sum operations that together produce a single output frame ($N$ values). This output corresponds to the ``polyphase'' step (the inner sum in \eqref{eq:pfb}), which can be understood by considering the operations acting on \textit{one} of the $N$ independent columns. Consider column $n$ of the data array, which contains timestream samples $x_n, x_{n+N}, x_{n+2N}, \dots$ etc.  These samples are drawn from the input timestream starting at sample $n$ and sub-sampled by a factor of $N$. Similarly, column $n$ of the window array is the $N$-fold sub-sampled window function. The output corresponding to column $n$ at every $\mathcal{SW}$ application (once every $N$ samples) is then simply the correlation\footnote{Correlation with the sub-sampled window function is equivalent to convolution with the full-rate filter described by \eqref{eq:chan_tod}. The window coefficients are the time-reversed filter coefficients; see \eqref{eq:chan_dft} for example.} of the sub-sampled timestream with the corresponding sub-sampled filter. Filtered outputs from all $N$ sub-filters are then passed to the Fourier transform block. In summary, in a PFB, $N$ sub-sampled timestreams are filtered by their respective $N$-distinct $p$-sized filters in parallel. In the remainder of this text, a sub-sampled filter is referred to as a \textit{sub-filter}, and a sub-sampled timestream as a \textit{sub-stream}.

As a concrete example, consider a $N$-point PFB with $p=4$ taps acting on a long timestream of length $bN$ ($b \gg p$). The output of sub-filter path $n$ can be written as  
\begin{equation}
    \mathbf{d}_n = W_n \mathbf{x}_{n},
    \label{eq: pfbsubsampled}
\end{equation}
where $\mathbf{x}_{n}$ is the sub-sampled input timestream corresponding to column $n$ and the operator $W_n$ is rectangular matrix of shape $[b-3, b]$ given by
\begin{equation}
\setlength{\arraycolsep}{2pt}
W_n =
\begin{bmatrix}
w_n     & w_{n+N} & w_{n+2N} &  w_{n+3N}  &\cdots & 0\\
0     & w_{n} & w_{n+N} &  w_{n+2N} & \cdots & 0 \\
\vdots  & \vdots  & \vdots  & \vdots & \ddots & \vdots \\
0 & \cdots &  w_{n} &  w_{n+N} & w_{n+2N} & w_{n+3N} \\
\end{bmatrix}
\label{eq: pfbsubproblem}
\end{equation}
The Toeplitz structure of operator $W_n$ clearly demonstrates the correlation of the sub-sampled timestream of path $n$ with a filter of size $p=4$.  By decomposing a PFB into $N$ independent and identical sub-problems, the task of inverting the PFB is reduced to that of inverting one sub-problem. We exploit this decomposition to develop a fast inversion strategy discussed in the next section.

\section{Inverse Polyphase Filter Banks}\label{sec: inverse pfb}

PFB inversion begins with applying the inverse FFT (IFFT) to the recorded PFB spectra. This step undoes the action of $\mathcal{F}$, leaving the task of decorrelating the output of the stacked $\mathcal{SW}$ operator, i.e., solving for $\mathbf{x}_n$ in \eqref{eq: pfbsubsampled} for all $N$ independent sub-filters. The system of equations given by \eqref{eq: pfbsubsampled} is underdetermined since there are more input values than output values. One way to solve this system is to use the Moore-Penrose pseudo-inverse, e.g., as explored by \citet{myrepo}. However, in practice, the PFB output often contains additional noise introduced by quantization, and this noise is amplified when a naive pseudo-inverse is applied without any additional regularization (details in \S\ref{sec: quantization noise}).  This section describes a method for solving \eqref{eq: pfbsubsampled} using circulant boundary conditions.  This method enables a fast inverse PFB (IPFB) implementation and elucidates the conditions under which quantization noise dominates.  (Methods for suppressing this noise will be addressed in~\S\ref{sec: mitigating quantization}.)
% Here, we adopt circulant boundary conditions for solving \eqref{eq: pfbsubsampled}. The resulting algorithm lets us understand where the quantization noise dominates, suggests methods to suppress the noise amplification, and permits a fast implementation that makes IPFB of large data volumes tractable.

\subsection{Circulant inversion}
Suppose $b - p + 1$ PFB frames were recorded using $b$ frames of input data according to \eqref{eq: pfbsubsampled}. As $b \gg p$, the matrix $W_n$ is nearly circulant, or equivalently, the correlation in \eqref{eq: pfbsubsampled} is nearly a circular correlation. By appending $p-1$ rows to $W_n$ (three additional rows for $p=4$), it can be promoted to a circulant matrix $W_n^{\mathrm{circ}}$ of size $b$.  Equation~\eqref{eq: pfbsubsampled} then becomes a circulant correlation,
\begin{equation}
    \mathbf{d}^\mathrm{circ}_n = W_n^{\mathrm{circ}}\mathbf{x}_n
                 = \mathbf{w}_n \circledast \mathbf{x}_n,
    \label{eq: circ_convolution}
\end{equation}
where $\circledast$ denotes a circular correlation, $\mathbf{d}^\mathrm{circ}_n$ is the original data $\mathbf{d}_n$ extended by $p-1$ entries corresponding to the circulant output, and the kernel $\mathbf{w}_n$ is the first row of $W_n$. The advantage of adopting this circulant model is that $W_n^{\mathrm{circ}}$ is diagonalized by the DFT operator, thus allowing FFTs to be used for decorrelation.  To implement circulant decorrelation in practice, we  extend the size of $\mathbf{d}_n$ such that it contains $b$ frames of real data, i.e., $\dim(\mathbf{d}_n) = \dim(\mathbf{d}^\mathrm{circ}_n) = b$. An estimate of the original timestream is then formed with the following procedure. Let $F$ denote a DFT of size $b$, and $F^\dagger$ its inverse. By the correlation theorem, 
\begin{equation}
    \hat{\mathbf{x}}_n
    = F^\dagger\!\left[\frac{F \mathbf{d}_n}{(F \mathbf{w}_n)^*} \right]
    = F^\dagger\!\left[\frac{1}{(F \mathbf{w}_n)^*}\right] \circledast \mathbf{d}_n,
    \label{eq: decorrelation}
\end{equation}
where $\hat{\mathbf{x}}_n$ denotes an estimate of $\mathbf{x}_n$, the asterisk ($*$) denotes an element-wise conjugate, and all multiplication and division operations in \eqref{eq: decorrelation} are also applied element-wise. 

As real data are not circulant, the last $p-1$ entries of the extended $\mathbf{d}_n$ differ from $\mathbf{d}^\mathrm{circ}_n$.  By enforcing circularity during decorrelation, the samples of the reconstructed $\hat{\mathbf{x}}_n$ are corrupted near the array boundaries.  However, the error in the circulant model approaches zero for samples that are sufficiently far from the edges (explained in more detail in Fig.~\ref{fig: convolution kernels}).  In use cases with high data rates, as is typical for radio astronomy, the loss from edge corruption is negligibly small even for modest lengths of data.
% Since typical radio astronomy use-cases use thousands to millions of frames per second, the fraction of data lost due to edge corruption is very small for even modest lengths of data.

In \eqref{eq: decorrelation}, the Fourier transform $F\mathbf{w}_n$ of the sub-sampled window is the frequency response of the $n$th sub-filter, with Nyquist frequency $B/N$. If the response of a sub-filter reaches zero, the reconstruction of the corresponding sub-stream $\mathbf{x}_n$ is ill defined at that frequency since inverting the correlation requires dividing by zero. In situations where $F\mathbf{w}_n$ is sufficiently small, round-off error and quantization noise in $\mathbf{d}_n$ dominate the recovered $\mathbf{x}_n$.

The entries of $F\mathbf{w}_n$ are also the eigenvalues of the circulant matrix $W_n^{\text{circ}}$. Since $W_n$ is obtained from $W_n^{\text{circ}}$ by deleting its last $p-1$ rows, the singular values of $W_n$ are well approximated by the eigenvalue magnitudes $|F\mathbf{w}_n|$. A vanishing eigenvalue therefore corresponds to a vanishing singular value, and any decorrelation algorithm suffers noise amplification along that mode. We accordingly use the terms \textit{eigenvalue spectrum} and \textit{frequency spectrum} of a sub-filter interchangeably.

\begin{figure}[tb]
    \centering
    \includegraphics[width=1.0\linewidth]{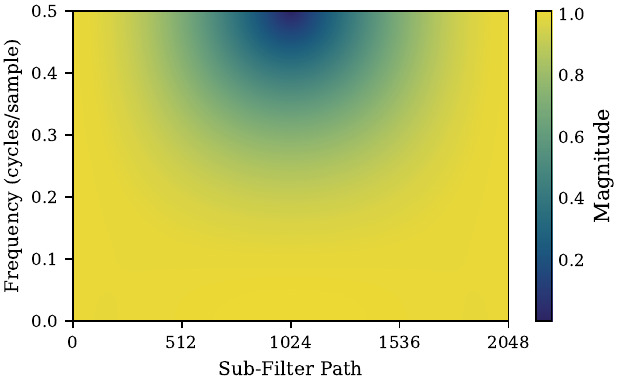}
    \caption{Magnitudes of the sub-filter eigenvalues for a $p=4$ tap, $N=2048$ point PFB with a sinc--Hamming window. Column $n$ shows $|F\mathbf{w}_n|$, the frequency spectrum of sub-filter $n$ (positive frequencies only, increasing upward); these are the eigenvalue magnitudes of the circulant operators $W_n^{\text{circ}}$ of Eq.~\eqref{eq: circ_convolution}. The normalized Nyquist frequency corresponds to a physical frequency of $B/N$, the width of one PFB channel. For sub-filters near $n = N/2$, the spectrum approaches zero toward the band edge (top center). Reconstruction in this region is degenerate: quantization noise dominates the decorrelation, limiting invertibility.}
    \label{fig: eigenvalues} 
\end{figure}

Invertibility of the PFB is assessed by examining the spectra of all $N$ sub-filters, shown in Fig.~\ref{fig: eigenvalues} for a PFB with $N=2048$, $p=4$, and a sinc--Hamming window. Each of the 2048 columns of Fig.~\ref{fig: eigenvalues} is the FFT of the zero-padded kernel $\mathbf{w}_n$ (positive frequencies only), i.e., the frequency spectrum of that sub-filter. The spectra of sub-filter paths near $n = N/2$ (center of the plot) approach zero at the band edge, suppressing those frequency modes of the corresponding sub-sampled timestream. Because each channel of the final PFB spectrum depends on the summed output of \textit{all} sub-filters, information lost in any one sub-filter prevents perfect reconstruction of the underlying spectrum when the inverted data are re-channelized at finer resolution. Appendix~\ref{sec: eigenvalues of symmetric pfb window} gives details of how this degeneracy is unavoidable for critically sampled PFBs with any symmetric window function.

\subsection{Quantization Effects}\label{sec: quantization noise}

% Quantization induces spurious noise spikes in the time-domain
Raw PFB data from an FPGA are often quantized using a limited number of bits because of constraints on data volume or bandwidth. Quantization introduces noise to the recorded spectra, and this noise is amplified during the PFB inversion process because of singularities in the convolution kernel. Circular decorrelation enables straightforward quantitative characterization of this corruption.

If the PFB output in \eqref{eq: pfboperator} is quantized using a high number of bits ($\gtrsim 4$), then under the Additive Quantization Noise Model \citep{fletcher2007robust, demir2020bussgang}, the recorded spectrum is $\mathcal{FSW}\mathbf{x} + \mathbf{n_q}$, where $\mathbf{n_q}$ is the quantization noise (difference between each sample in $\mathcal{FSW}\mathbf{x}$ and the nearest quantized level.)  As different values of $\mathbf{n_q}$ are uncorrelated with each other and are approximately uniformly distributed, $\mathbf{n_q}$ can be treated as white noise when assessing its impact on the IPFB.

After inverting $\mathcal{F}$, the decorrelation of $\mathcal{SW}$ is performed by dividing the DFTs of the sub-filters by the conjugate DFT of $\mathrm{W}_n$, but in the presence of noise:
%Noisy decorrelated estimate of the timestream can be written as
\begin{equation}
    \hat{\mathbf{x}}_n = F^\dag\left\{\frac{F \mathbf{d}_n}{(F \mathbf{w}_n)^*} + \frac{F \boldsymbol{\epsilon}_n}{(F \mathbf{w}_n)^*}\right\}.
    \label{eq: noisy_deconv}
\end{equation}
Here $\boldsymbol{\epsilon}_n$ is the time-domain version of the original quantization noise $\mathbf{n_q}$, obtained after application of $N$-point IFFT to the PFB spectra, i.e. $\boldsymbol{\epsilon}_n = (\mathcal{F}^{-1} \mathbf{n_q})_n$. For sub-filter paths close to $N/2$, $F\mathbf{w}_n$ is nearly singular, causing the quantization noise term in~\eqref{eq: noisy_deconv} to be strongly amplified during reconstruction. Consequently, the reconstruction error, $\hat{\mathbf{x}}_n-\mathbf{x}_n$, in samples corresponding to these sub-filter paths is much higher than would be naively expected from the quantization noise in the PFB data. For example, when a PFB with $N=2048$-point DFTs is inverted, the RMS reconstruction error in samples corresponding to sub-filter path $N/2=1024$ is many times higher than in samples from other filter paths. The reconstruction error therefore appears in the timestream as periodic noise spikes concentrated near the center of each frame. When the IPFB timestream is channelized again at higher frequency resolution, these periodic timestream spikes produce a corresponding picket-fence pattern of noise spikes in the frequency spectrum.

This decorrelation operation can be equivalently viewed as correlation with a kernel having a frequency response of $1/F \mathbf{w}_n$. Examining the time-domain response of this ``inverse" kernel for various sub-filter paths provides an alternate approach for understanding the noise amplification in the reconstructed timestream. Decorrelation kernels for two sub-filter paths are contrasted in Fig.~\ref{fig: convolution kernels}. Plots labeled ``unfiltered" correspond to inverse kernels used for decorrelation in ~\eqref{eq: noisy_deconv}, and are discussed in this section. Since the inverse kernel for path 1024 has a wide region of support, one sample of reconstructed sub-stream $\hat{\mathbf{x}}_{N/2}$ receives noise contributions from several thousand samples around it, thereby boosting the noise floor. Inverse kernels for sub-filter paths far away from $N/2$, however, approach a delta function. Therefore, a sample of sub-stream $\hat{\mathbf{x}}_{0}$, for example, contains quantization noise from that sample alone.

The qualitative behaviour of reconstruction error in the presence of quantization noise described here will be shown to agree with detailed numerical simulations in the next section, which describes various methods to mitigate the effects of quantization noise.

\section{Mitigating Quantization Effects}\label{sec: mitigating quantization}

This section describes and compares two noise mitigation methods. The first is a Wiener filter approach, which employs prior knowledge of the noise and signal statistical properties and can be applied because the noise is additive.  The second is a maximum-likelihood optimization formulation of the IPFB, in which a time-domain prior containing information on the degenerate regions of the filter response is imposed to improve the conditioning of the correlation matrices.

\subsection{Wiener Filter}\label{sec: wiener filter}

\begin{figure}[tb]
    \centering
    \includegraphics[width=1.0\linewidth]{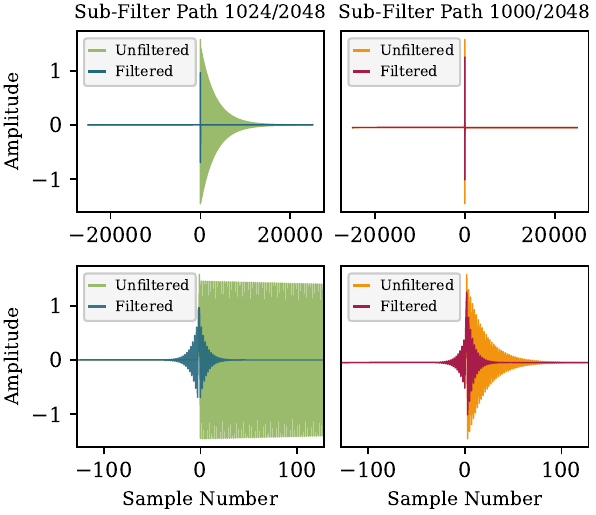}
    \caption{Time-domain representation of IPFB decorrelation kernels for a 2048-point, 4-tap PFB that uses a sinc-Hamming window function. Decorrelation kernels are obtained as the FFT of the inverse filter response, $H_n/(F\mathbf{w}_n)^*$. In this example, $H_n$ is either 1 (naive decorrelation) or the Wiener filter, discussed in section~\ref{sec: mitigating quantization}. The two columns of panels show two different sub-filter paths. The bottom panels are zoomed-in versions of the top panels. Samples on the horizontal axis correspond to a sampling rate of $2B/N$ (raw timestream sub-sampled by factor of $N$). The kernels show that sub-filters near the frame center (1024) have a broad but finite domain of support: the worst-conditioned filter path decays to 0 in approximately 20,000 samples. As the sub-filter path moves away from the center, the corresponding kernel approaches a delta function. Therefore, in the absence of quantization noise, circulant decorrelation approaches floating-point precision away from the edges of the recovered timestream. Upon decorrelation, the 3-sample error mentioned in section~\ref{sec: inverse pfb} corrupts only as many samples as the width of the worst-conditioned filter path. When inverting large volumes of PFB data,
    %(in radio astronomy, one second of data is typically tens of thousands of frames)
    the fraction lost to edge-corruption is insignificant.
    }
    \label{fig: convolution kernels}
\end{figure}

\begin{figure}[tb]
    \centering
    \includegraphics[width=\linewidth]{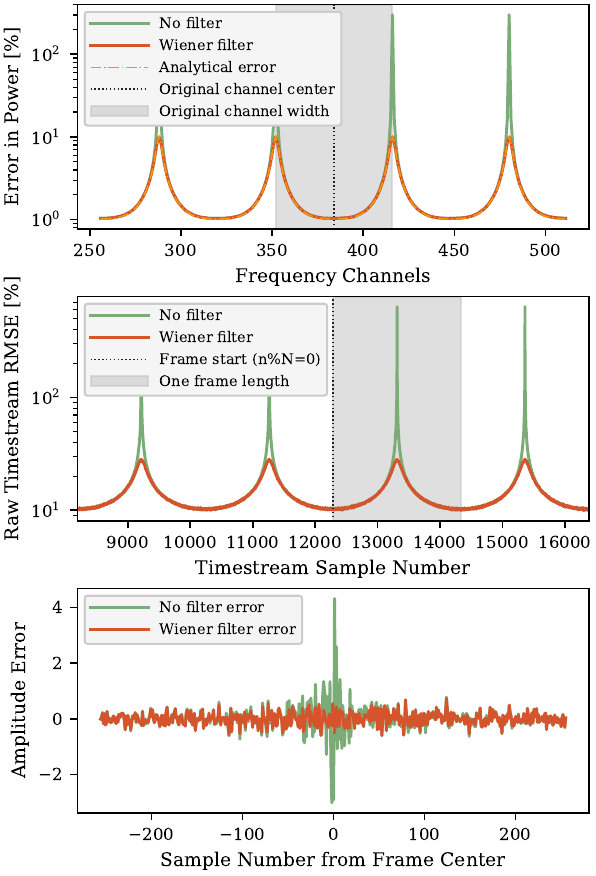}
    \caption{Numerically simulated comparison of percentage error in time-stream reconstruction and re-channelized power spectrum. The simulation PFB'd a white-noise timestream, quantized the spectra to 4-bits, and inverted the spectra using the two shown methods. The reconstructed timestream re-channelized using 64-times longer frame sizes for finer frequency resolution. Top plot shows percentage error in the frequency spectrum of re-channelized spectrum compared to channelization of the original timestream, averaged over many spectra. Middle plot shows reconstruction error in the timestream as compared to the original timestream. It can be seen that samples near frame centre are corrupted. Bottom plot zooms-in on a frame of filtered and unfiltered IPFB timestreams to highlight the different levels of reconstruction noise. The channels with 10\% error can either be discarded, or the number of quantization bits can be increased, which will rapidly improve the re-construction error. (See appendix \ref{sec: optimal wiener threshold}.)} %well that's only 4 bits though. I'll mention that bias goes down with more bits.
    \label{fig: wiener comparison}
\end{figure}

When the covariance, or equivalently the power spectral density (PSD) of the signal and noise are known, the Wiener filter yields the optimal\footnote{in a least-squares sense.} estimate of the underlying signal from a noisy observation \citep{wiener1949extrapolation, rybicki1992interpolation}. The general form of the filter is $S/(S+N)$, where $S$ and $N$ are, respectively, the PSD of signal and noise. For the case of inverting a PFB, the appropriate form of the filter for decorrelating the effect of the sub-sampled window function from sub-filter path $n$ is
\begin{equation}\label{eq: wiener weights}
     H_n(f)= \frac{|W_n(f)|^2}{|W_n(f)|^2 + \phi^2} (1 + \phi^2),
\end{equation}
where $W_n(f) = F \mathbf{w}_n$ is the frequency spectrum of sub-filter $n$ (vertical slices in Fig.~\ref{fig: eigenvalues}), and $\phi$ is a threshold parameter that controls the strength of noise suppression. The multiplicative factor $1+\phi^2$ is a normalization term that keeps power unchanged where $W_n = 1$. Fig.~\ref{fig: wiener comparison} shows the effects of Wiener filtering on reducing the percentage RMS error on the recovered time-stream samples, and on the auto-spectrum of the re-channelized PFB data. The fractional error on auto-spectrum is calculated as
\begin{equation}
    \delta(f) = \frac{\langle |X(f) - \hat{X}(f)|^2 \rangle}{\langle |X(f)|^2 \rangle},
    \label{eq: autospectrum_error}
\end{equation}
where $X(f)$ and $\hat{X}(f)$ are the high-frequency-resolution spectra of the original timestream and the IPFB timestream, respectively. The ensemble average is taken by averaging many spectra along the time axis (described in more detail in section \ref{sec: quantization-mitigation-results}). Physically, $\phi$ is related to the quantization SNR\footnote{defined as the ratio of signal variance to quantization noise variance} as $\phi^2 = 1/\mathrm{SNR}$.  Numerical simulations show that, as a rule of thumb, $\phi \sim 0.1$ generally provides sufficient noise suppression for 4-bit-quantized PFB data. 

The periodic nature of time- and frequency-domain noise spikes described in the previous section is apparent from  the simulation results in Fig.~\ref{fig: wiener comparison}.  Both the time- and frequency-domain spikes are broadened because inversion problems arise from a range of sub-filter eigenvalues that lie near zero.
%Notice, however, that both time-domain and frequency-domain spikes are not sharp. The spikes are wide because the eigenvalues of a sub-filter do not need to be exactly zero to cause inversion problems.
From \eqref{eq: noisy_deconv}, invertibility issues are expected when the response of any sub-filter drops (the region near top-center in Fig.~\ref{fig: eigenvalues}) below the threshold set by the quantization noise. Consequently, corruption appears in small regions near a frame center and near the boundaries of the original channels in the re-channelized spectrum. By Wiener filtering, power in the low SNR modes is suppressed, preventing noise amplification. Equivalently, from the time-domain decorrelation viewpoint, a Wiener filter shrinks and restricts the region of support of inverse kernels (plots labelled ``filtered" kernels in Fig.~\ref{fig: convolution kernels}).  The exact form of the Wiener filter, the optimal value of $\phi$, and an analytical expression for $\delta(f)$ are derived in appendix~\ref{sec: optimal wiener threshold}. The computational cost of the Wiener-filtered IPFB algorithm is compared with that of other filtering-based IPFB approaches in appendix~\ref{sec: time complexity}.

The Wiener filter is one statistical approach to down-weighting noisy eigenmodes. More generally, because the circulant formalism explicitly provides the eigenspectrum of the PFB transform matrix, arbitrary eigenvalue filtering schemes can be applied at negligible additional computational cost. 

\subsection{Maximum-Likelihood Optimization}\label{sec: salvaging the inverse}\label{sec: conjugate gradient}

\subsubsection{Time Domain Prior}
An alternate approach to spectral filtering is formulating timestream reconstruction as a maximum likelihood (ML) estimation problem. While more computationally intensive, ML-method can improve upon the Wiener filter results. Given some data $d$, a linear model $A$, model parameters $x$, and a covariant, positive-definite noise matrix $N$, we write
\begin{equation}\label{eq: chi squared without prior}
    \chi^2 = (d - Ax)^TN^{-1}(d-Ax).
\end{equation}
Minimizing $\chi^2$ yields the model parameters most likely to have produced the observed data. For the case of computing an IPFB, $d$ is the quantized, channelized data after the PFB is applied, $x$ is the unknown time-domain data, and $A$ is the PFB operator given by \eqref{eq: pfboperator}. Quantization noise is assumed to be uniform and uncorrelated such that the data noise covariance matrix, $N$, is proportional to the identity matrix. This least-squares system can be solved using standard techniques like conjugate gradient (CG) \citep{shewchuk1994introduction} to obtain an estimate of the original timestream.  In practice, an inverse-FFT (IFFT) of the PFB data is taken (undoing $\mathcal{F}$) before solving the system, such that $A=\mathcal{SW}$, which makes the $\chi^2$ evaluation faster.  

A direct least-squares solution to \eqref{eq: chi squared without prior} inverts the near-zero singular values of the $\mathcal{SW}$ matrix, which, again, leads to the undesirable noise amplification. To better constrain the problem, a prior/regularization term must be included to fill in the attenuated singular modes. In 
\S\ref{sec: wiener filter}, prior information was used by specifying the PSD of the signal and the noise (assumed to be known a-priori), which led to the Wiener filter formalism. Alternately, the requirement of known PSDs can be sidestepped by specifying a time-domain prior. Since reconstruction noise is highest near the centre of a PFB frame, if a subset of the original timestream samples corresponding to the corrupted sub-filter paths are retained, they can be incorporated as a prior in the ML optimization as follows:
\begin{equation}\label{eq: chi squared}
    \chi^2 = (d - Ax)^T N^{-1} (d - Ax) + (x - p)^TQ^{-1}(x - p).
\end{equation}
The pre-saved time-stream samples are captured by the vector $p$, and the inverse noise matrix in the prior term, $Q^{-1}$, is 1 for time indices saved and 0 for indices discarded. The prior term then evaluates to $\sum_i (x_i-p_i)^2$, where $i$ iterates over the saved timestream samples.
Practical implementation of this approach requires modifying the digital logic for conditionally re-routing a fraction of raw timestream samples. Consequently, the choice of which samples to retain from each sub-filter path directly determines the additional digital bandwidth required by the hardware for enabling this algorithm. 

% In the next section, we quantitatively compare the two methods for PFB inversion using simulated data, and investigate if the improvement in PFB inversion fidelity with the time-domain prior is worth the added computational complexity.

\subsubsection{Implementing the Prior}
Consider a toy model of an $N$-point, $p$-tap PFB that uses a symmetric window and has no output quantization. For this PFB, the sub-sampled window array for filter path $N/2$, $\mathbf{w}_{N/2}$, is of the form $[a,b,b,a,0,0,\ldots]^T$. Due to symmetry, the eigenvalue spectrum of this filter path is exactly zero at the Nyquist frequency (see appendix \ref{sec: eigenvalues of symmetric pfb window}). The correlation operation of the PFB, given by~\eqref{eq: pfbsubsampled}, therefore loses all information about the Nyquist mode of the sub-sampled timestream $\mathbf{x}_{N/2}$. Decorrelating the Nyquist mode $\mathbf{x}_{N/2}$ is now ill-defined because it involves division by a zero eigenvalue. All other eigenmodes, however, can be successfully recovered. 

Suppose we now take a pseudo-inverse, where the offending zero eigenvalue is truncated. The reconstructed $\hat{\mathbf{x}}_{N/2}$ then differs from the original $\mathbf{x}_{N/2}$ only in the Nyquist frequency. In particular, the residual is $\mathbf{r} = \alpha [-1,1,-1,1,\ldots]^T$, for some $\alpha \in \mathbb{R}.$ Moreover, $\mathbf{r}$ lies in the null-space of $W_{N/2}$, so that moving in that direction of makes no difference in $\chi^2$ given by~\eqref{eq: chi squared}. This missing information (or degenerate direction) can be constrained with a prior.
Including prior information in this example is straightforward. If we save just a single sample of the original timestream, we know $\alpha$, which in turn means we know $\mathbf{r}$.  Subtracting $\mathbf{r}$ from our reconstructed $\mathbf{x}_{N/2}$ then gives us the true $\mathbf{x}_{N/2}$.
%If one sample from the original $\mathbf{x}_{N/2}$ is saved, enough to infer one entry of $\mathbf{r}$, all entries of $\mathbf{r}$ are trivially calculated, which permits perfect reconstruction of $\mathbf{x}_{N/2}$. 

\begin{figure}[tb]
    \centering
    \includegraphics[width=\linewidth]{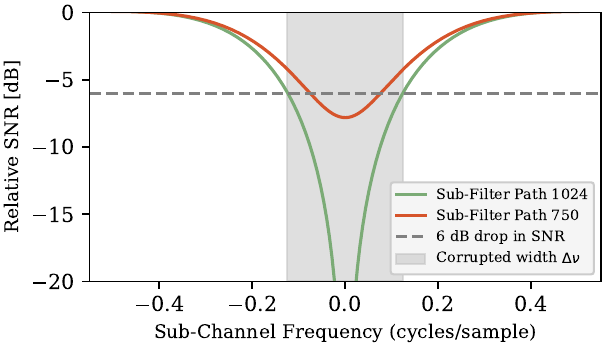}
    \caption{SNR (in dB) relative to the maximum possible SNR that can be obtained upon a naive ``no filter" IPFB. For a given quantization scheme, the relative SNR is simply by $1/|F\mathbf{w}_n|^2$. The abscissa spans the width of one original frequency channel. The two curves show the SNR behaviour for two different sub-filter paths for a 2048-point PFB. The width of the frequency band where SNR drops more than 6 dB is marked as corrupt, and used to guide the sample saving strategy for time-domain prior method described in section~\ref{sec: conjugate gradient}. Gray highlighted region illustrates corrupted region for the filter path $N/2$. Filter paths away from $N/2$ have a smaller corrupted bandwidth.}
    \label{fig: corrupted_region}
\end{figure}

In a real PFB, the introduction of quantization noise corrupts a small but finite band of frequencies as shown in Fig.~\ref{fig: wiener comparison}. In this case, we need to save more than just one sample, but still only a tiny fraction of total samples. If the width of the corrupted band is $\Delta \nu$, the correlation length of the resulting residual timestream is of the order $1/\Delta \nu$. Thus, samples of $\mathbf{x}_{N/2}$ saved at a rate of roughly $\Delta \nu$ contain sufficient information to enable reconstruction of the corrupted modes. Instead of taking a pseudo-inverse and then solving for the missing modes, as in the toy example, the saved samples can now be used as a principled prior in the full-least squares solution to~\eqref{eq: chi squared}.

If one out of every $R$ samples is saved from $L$ out of $N$ sub-filter paths, the fraction of the original timestream saved is
\begin{equation}
    \beta = \frac{L}{RN}
\end{equation}
In our numerical experiments (described in the following section), we fix $R=4$, i.e. save every 4th time-sample of a channel, and vary $L$ to obtain the desired level of saved fraction, $\beta$. For guiding the eye, the choice of $R=4$ corresponds to a corrupted bandwidth of approximately 20\% for the sub-filter path $N/2$, as depicted in Fig.~\ref{fig: corrupted_region}. We emphasize, however, that $R$ and $L$ are tunable parameters, and other, more efficient sample saving strategies are possible. For example, filter paths away from $N/2$ have lower levels of noise corruption (smaller $\Delta \nu$), such that their samples  can be saved at a slower rate.

\subsection{Quantization Mitigation Results}\label{sec: quantization-mitigation-results}

\begin{figure}[tb]
    \centering
    \includegraphics[width=\linewidth]{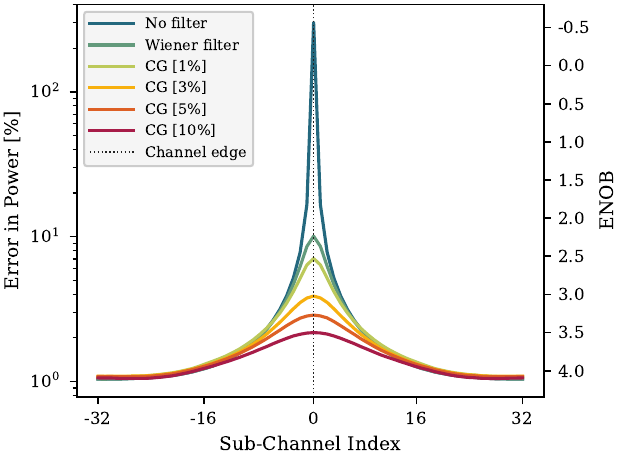}
    \caption{Percentage error in re-channelized autospectrum for varying levels of time-domain samples saved. The simulation used here is identical to the one used for the Wiener-filtering results shown in Fig.~\ref{fig: wiener comparison}, and described in detail in section~\ref{sec: quantization-mitigation-results}. For this plot, simulated 4-bit-quantized PFB data with 2048 channels was inverted with a conjugate gradient (CG)-based maximum likelihood optimizer, where the fraction of samples retained from the original timestream was varied from 1\% to 10\%. The IPFB timestream was re-channelized with 64-times finer channel resolution. The displayed bandwidth is one original channel or 64 fine channels wide. The plots are centered at the edge of two original channels to highlight the reconstruction error. With only 3\% of timestream samples saved, the worst-case error drops to less than half of Wiener-filter error. Saving 10\% of timestream samples further brings the error down to 2\%. For comparison, the noise floor set by 4-bit quantization is approximately 1\%. The secondary y-axis shows the effective number of bits (ENOB) obtained from effective quantization SNR (1/error) using the relation in~\eqref{eq: snr_enob}.}
    \label{fig: cgerror}
\end{figure}

The effects of quantization noise on the reconstructed timestream and the re-channelized spectrum, obtained using the two IPFB methods, are assessed using an end-to-end simulation consisting of the following steps. A digital timestream of length 2\textsuperscript{27} is generated from the standard normal distribution to simulate a signal incident upon an antenna. This timestream is passed through a 2048-point, 4-tap PFB, and the real/imaginary components of the resulting complex spectra are quantized using a 15-level (4 bits), mid-tread quantizer (consisting of uniformly-spaced levels from -7 to 7, with a 0 level.)  The level-spacing for the quantizer was chosen to be 0.353 of the RMS of the channelized spectra as this value minimizes the mean-squared quantization error for a 4-bit uniform quantizer\citep{max1960quantizing}. The PFB is then inverted in three ways: circulant inversion without filtering, circulant inversion with Wiener filtering, and CG inversion with varying fractions of raw timestream samples saved. Finally, the reconstructed timestream is channelized again using a PFB with 64-times longer frame size, to produce 64-times finer frequency resolution. The fractional error in the rechannelized autospectrum, denoted by $\delta(f)$, is calculated using~\eqref{eq: autospectrum_error}, where the true spectra is obtained by directly channelizing the original timestream to the same, 64-times higher frequency resolution. As $\delta(f)$ is the ratio of jitter in power relative to the true power, the metric $1/\delta(f)$ indicates the effective SNR achieved in a channel after rechannelization. To assess the quality of raw timestream reconstruction, fractional error in timestream is calculated as $\sqrt{\mathrm{Var}[x - \hat{x}]}/\sigma_x$, where $\sigma_x$ is the standard deviation of the original timestream. Both error metrics are estimated by averaging several thousand PFB frames. 

Error in rechannelized autospectrum obtained after naive and Wiener filter IPFB are compared in the top plot of Fig.~\ref{fig: wiener comparison}. The result shows that near the center of an original channel, where all correlation kernels are well-behaved, $\delta(f)$ is approximately 1.1\%. In units of decibels, the relative error is $10\mathrm{log_{10}}[\delta(f)] \approx -19\; \mathrm{dB}$, which is consistent with the SNR ceiling of approximately 19 dB set by our 4-bit quantization scheme~\citep{hui2001asymptotic}. Near the channel edges, where $F\mathbf{w}_n$ drops precipitously for sub-filter paths close to $N/2$, the Wiener filter is seen to suppress noise amplification. With Wiener filtering, the maximum value of $\delta(f)$ near the channel edges is approximately 10\%, with the worst-affected region occupying less than 10\% of the channel width. This error level also corresponds to salvaging 2.5 bits of information near the channel edges (see Fig.~\ref{fig: cgerror}). In comparison, the worst-case error is more than 100\% without filtering, resulting in near-total loss of information for some sub-filter paths.

The fractional RMS error in timestream obtained using naive and Wiener filter IPFB is compared in the middle plot of Fig.~\ref{fig: wiener comparison}.  In agreement with the autospectra results, Wiener filter decorrelation limits the noise amplification for samples near the center of the original PFB frame (highlighted in gray), which correspond to near-singular sub-filter paths. Most of the PFB frame samples have errors of about of 10\%, which is the 4-bit quantization noise floor in amplitude.

Results obtained with CG $\chi^2$ minimization are displayed in Fig.~\ref{fig: cgerror}.  The autospectrum error, $\delta(f)$, is shown for four levels of saved fraction, $\beta$, along with the Wiener filter and no-filter results for comparison. For each $\beta$, the CG solver was provided the Wiener filter solution as the initial guess and was run for 25 iterations. The rechannelization error is seen to reduce smoothly, and approaches the quantization noise floor, as samples from more sub-filter paths are included in the prior term. For $L=4$ and $N=2048$, an increase in $\beta$ from 1\% to 10\% corresponds to an increase in the number of saved paths from 82 to 820 (symmetric around center of the frame, 1024), where every 4th sample from each path is retained. Near the edges of the channel, the CG error is seen to drop to less than half of the Wiener filter solution with retention of only 3\% of raw timestream samples. Equivalently, 3 out of 4 bits of information can be recovered in the worst-corrupted region by retaining 3\% of timestream samples. With 10\% of samples saved, the error reaches approximately 2\%, or within 3 dB of the 4-bit quantization floor of approx. 1\%. 

To put the additional resource demand of ML method in perspective, the floating-point oversampled-PFB reconstruction reported by~\citealp{morrison2020performance} usex an oversampling factor of 4/3, which requires 33\% higher digital bandwidth than an equivalent critically-sampled PFB. For quantized PFBs, our sparse time-domain prior method achieves reconstruction to within 6 dB of the maximum achievable SNR with 4-bit quantization with only 3–5\% of samples saved, which is a substantially lower requirement on additional bandwidth. We expect that more sophisticated sample saving schemes might enable comparable results with an even smaller fraction of samples saved, but we defer a full exploration of these schemes to future work.

\section{Conclusion}

This paper presents an efficient, Fourier-transform-based algorithm to invert critically sampled PFB measurements, enabling offline ``up-channelization'' of data to finer frequency resolution. By imposing circulant boundary conditions, the PFB inversion can be expressed as a set of independent decorrelations that are diagonal in the Fourier basis, providing a direct path to an efficient FFT-based implementation. This formalism also exposes a fundamental limitation: the problem is intrinsically ill-conditioned at frequencies where a polyphase sub-filter's response approaches zero, and quantization noise is subsequently amplified strongly during inversion. The circulant formalism makes it straightforward to characterize the exact behaviour of this noise amplification in the time and frequency domains .

We explore two approaches for mitigating this noise amplification. The first applies a Wiener filter directly to the sub-filter eigenmodes in Fourier space, suppressing modes in which quantization noise dominates the reconstruction. For 4-bit quantized data, Wiener filtering reduces the worst re-channelized spectral errors from greater than 100\% to approximately 10\%, with significant errors confined to less than 10\% of the original channel width. The reconstruction fidelity improves rapidly with increasing quantization depth. The Fourier-domain implementation allows Wiener filtering to be applied with negligible additional computational cost.

The second approach casts PFB inversion as a maximum-likelihood optimization problem, using a sparse prior of retained time-domain samples to constrain poorly measured modes. In simulations of 4-bit quantized data, retaining only 3\% of the original timestream reduces the worst-case spectral error to less than half that obtained with Wiener filtering, and retaining 5–10\% of samples further brings down to within 3 dB of the 4-bit noise floor. Although this approach requires additional data acquisition and greater computational effort compared to the FFT-based Wiener filter method, only a small amount of complementary time-domain information is needed to substantially improve reconstruction fidelity. 

For conventionally recorded PFB data, the Wiener-filtered inverse provides a robust and computationally efficient approach to inversion/up-channelization while controlling the noise amplification inherent to the inverse problem. In a future paper, we will demonstrate the performance of a scalable GPU implementation of the Wiener filter IPFB algorithm, which can invert many days of data collected with the ALBATROS radio telescope in a continuous, streaming fashion. In applications where greater reconstruction fidelity is required, our maximum-likelihood results demonstrate that retaining only a small fraction of the original timestream can recover much of the information lost to poorly constrained modes. Together, these results provide a practical framework for understanding and mitigating the limitations of critically sampled PFB inversion, enabling finer spectral resolution to be recovered from channelized radio-astronomical data.

%% Please use the acknowledgment and contribution environments. This will 
%% be anonomyized when the "anonymous" style option is used. 
\begin{acknowledgments}
We acknowledge the support of the Natural Sciences and Engineering Research Council of Canada (NSERC), RGPIN-2019-04506, RGPNS 534549-19; Canada Foundation for Innovation John R. Evans Leaders Fund \#40824; National Geographic Society Explorer Grant NGS-94983T-22. We acknowledge the support of the Government of Canada’s New Frontiers in Research Fund (NFRF) NFRFE-2023-00069. This research was undertaken, in part, thanks to funding from the Canada 150 Research Chairs Program. MA acknowledges support from the Vanier Canada Graduate Scholarship Program.
\end{acknowledgments}

\begin{contribution}
%%This section gives authors the space to recognize author contributions. The text inside this environment is NOT counted towards the total word quanta. At a minimum, manuscripts are expected to include this text:

%%All authors contributed equally to the Terra Mater collaboration.

SF wrote the initial version of Wiener filter and CG solvers for inverse PFB, investigated their performance and behavior, created the visualizations, and wrote an initial draft of the manuscript.
MA conducted the literature review, performed formal mathematical analysis of the IPFB methods, analyzed and validated simulation results, and co-authored this manuscript (including the appendices). He also contributed to the development of IPFB methods and ideas for visualization.
HCC reviewed and edited the manuscript, provided feedback on the visualizations, and supervised the presentation of technical aspects of the IPFB methods.
AC contributed in creating and refining the visualizations, participated in technical discussions on IPFB methods, and collaborated on writing the manuscript.
JS conceived the initial research idea, contributed to the development of IPFB methods and interpretation of results, reviewed and edited the manuscript, and supervised the project from its conception to completion.
ST conducted initial investigation into possible approaches for inverting PFBs, reviewed the manuscript, and provided feedback.

%%
%% Authors can use the Contributor Role Taxonomy (CRediT) at
%% https://credit.niso.org
%% for ideas on how write a good statement tailored to their needs.

\end{contribution}

%% To help institutions obtain information on the effectiveness of their 
%% telescopes the AAS Journals has created a group of keywords for telescope 
%% facilities.
%
%% Following the acknowledgments section, use the following syntax and the
%% \facility{} or \facilities{} macros to list the keywords of facilities used 
%% in the research for the paper.  Each keyword is check against the master 
%% list during copy editing.  Individual instruments can be provided in 
%% parentheses, after the keyword, but they are not verified.
% \facilities{HST(STIS), Swift(XRT and UVOT), AAVSO, CTIO:1.3m, CTIO:1.5m, CXO}

%% Similar to \facility{}, there is the optional \software command to allow 
%% authors a place to specify which programs were used during the creation of 
%% the manuscript. Authors should list each code and include either a
%% citation or url to the code inside ()s when available.
\software{Numpy \citep{harris2020array},  
          Scipy \citep{2020SciPy-NMeth},
          Matplotlib \citep{hunter2007matplotlib}.
          }

%% Appendix material should be preceded with a single \appendix command.
%% There should be a \section command for each appendix. Mark appendix
%% subsections with the same markup you use in the main body of the paper.
%%
%% Each Appendix (indicated with \section) will be lettered A, B, C, etc.
%% The equation counter will reset when it encounters the \appendix
%% command and will number appendix equations (A1), (A2), etc. The
%% Figure and Table counter will not reset.

\appendix

\section{Eigenvalues of a symmetric PFB window}\label{sec: eigenvalues of symmetric pfb window}

The singularities encountered in sub-filter responses shown in Fig.~\ref{fig: eigenvalues} arise, fundamentally, due to symmetry properties of the window function used in the PFB. Here, we provide a heuristic proof of why any reasonable choice of window function will always lead to singular values in the critically-sampled PFB operator. We require a sensible window array $W$ of length $M$ to be: (1) symmetric about its center, such that $W_k=W_{M-k}$, and (2) go to zero on the boundary, so that $W_0=W_{M-1}=0$. The latter requirement is reasonable because windows are designed to suppress spectral leakage by gently approaching zero towards the boundaries of a finite timestream.

Consider a $N$-point, $p$-tap PFB with $M=pN$ constructed using $W$, and let $p$ be an even number. From \eqref{eq: circ_convolution}, a PFB correlates a sub-sampled input timestream with the sub-sampled window array, $\mathbf{w}_n$, which comprises only $p$ non-zero entries. For the sub-filter path $N/2$, these $p$ entries are exactly symmetric. For example, with $p=4$ taps, $\mathbf{w}_{N/2}$ is of the form $[a,b,b,a,0,0,\ldots]$ due to the symmetry requirement. To evaluate the the eigenvalues (or frequency response) of this sub-filter, a $M$-point Fourier transform, $F \mathbf{w}_n$, is performed. The eigenvalue corresponding to the Nyquist frequency is then seen to be a dot product of $[a,b,b,a,0,0,\ldots]$ with the alternating sequence $[1,-1,1,-1,\ldots]$; the latter being the DFT phasor $e^{j\pi m}$ for the Nyquist frequency. This dot-product evidently vanishes, corresponding to the point of singularity at the top-center of the plot in Fig.~\ref{fig: eigenvalues}. For an odd number of taps, it is the eigenvalue of path $0$, instead of the $N/2$, that vanishes. Because $W_0=0$, the dot product of remaining $p-1$ elements of $\mathbf{w}_{0}$ evaluates to zero due to symmetry\footnote{Consider a 5-tap PFB as a concrete example.  The window will be symmetric around sample $5N/2$ where $N$ is the number of samples per frame.  Path 0 will consist of samples at $[0,N,2N,3N,4N]$.  If the window goes to 0 at the edge then sample 0 will be zero.  The symmetry around $5N/2$ means the window at samples $2N$ and $3N$ are the same, and that the window at samples $N$ and $4N$ will be the same.  That leaves $\mathbf{w_0}=[0,a,b,b,a,0...]$, which will Fourier transform to zero at the Nyquist frequency.}.

The more closely a window adheres to the two requirements of symmetry and zero at the boundaries, the closer the smallest PFB eigenvalue is to zero. Consequently, inverse PFB suffers a correspondingly higher noise amplification. Common windows implemented in the \texttt{Numpy} package, such as Hamming, Hanning, and Blackman are not exactly symmetric in $M$ due to the normalization choice of $cos[2 \pi m / (M-1)]$, but they approach perfect symmetry with increasing $M$. The IPFB noise amplification problem is, therefore, worse for PFBs that use high FFT frame sizes, $N$, or too many taps, $p$.

\section{Optimal Wiener Threshold}\label{sec: optimal wiener threshold}

A radio telescope's signal processing chain quantizes the real and imaginary parts of the output of a PFB. In what follows, let $s$ be the real (or imaginary) component of a PFB channel and $\mathcal{Q}(s)$ its quantized version. In an optimally-tuned quantizer, all quantization levels are well-exercised and the clipped fraction is kept low. For such a quantizer, the RMS of quantization noise, $\sigma_q$, defined as $\sigma_q = \sqrt{\mathrm{Var}[s - \mathcal{Q}(s)]}$ is approximately $\Delta/\sqrt{12}$, where $\Delta$ is the level spacing or one least-signifcant bit (LSB) of the quantizer. For the example of 4-bit (15 level, symmetric) quantization used in section~\ref{sec: conjugate gradient}, the signal\footnote{by ``signal'' we mean antenna timestream. Noise refers specifically to quantization noise, and all references to SNR refer to quantization SNR in this section.} RMS $\sigma_s$ and $\Delta$ are related as $\Delta = 0.353 \sigma_s$ in the optimal case. The signal-to-noise ratio (SNR), defined as $\sigma_s^2/\sigma_q^2$, achieved with this tuning is approximately 80. In decibels, $\mathrm{SNR_{dB}} = 10\mathrm{log_{10}}[\delta(f)]$, the optimal SNR is 19 dB, and for a given signal level, it increases by approximately 6 dB per additional bit. Another useful metric assessing the SNR achieved by a given signal processing system is effective number of bits (ENOB), which quantifies SNR in terms of bits of information retained or lost. For example, an ENOB of 3 implies that the SNR of the system is equivalent to that achieved by an optimally-tuned 3 bit digitizer. To convert between quantization SNR and ENOB, we use the following relation
\begin{equation}
    \mathrm{SNR_{dB}} = 5.2 \mathrm{R} - 1.8,
    \label{eq: snr_enob}
\end{equation}
where, $R = log_2 N_{bits}$. This approximate scaling relation was obtained fitting a line to the results obtained for uniform quantization of a gaussian signal by~\citealp{hui2001asymptotic}, displayed in their figure 2. In \S\ref{sec: mitigating quantization} in the main text, ENOB is used to asses the amount of information successfully recovered by various IPFB methods in the worst-corrupted portions of the rechannelized spectrum.

After taking an N-point IFFT of the PFB spectra, the decorrelation problem for a sub-filter path $n$ ($n < N$) is $\mathbf{d}_n = W_n^{\mathrm{circ}}\mathbf{x}_n + \boldsymbol{\epsilon}_n$, where $\boldsymbol{\epsilon}_n$ is the original quantization noise scaled by the IFFT and window normalization factors. Since the scaling factors are common to both the quantization noise, $\boldsymbol{\epsilon}_n$, and the signal, $\mathbf{x}_n$, the SNR in time-domain does not change. At this stage, Wiener filtering $\mathbf{d}_n$ would suppress power in the noisy eigenmodes, preventing noise amplification upon decorrelation. To obtain the exact expression in~\eqref{eq: wiener weights}, let the signal power spectral density (PSD) be $S(f) = |W_n(f)|^2 \langle |X_n(f)|^2\rangle$, and the noise PSD be  $N(f) = \langle |Q_n(f)|^2\rangle$. Here $W_n(f)$, $X_n(f)$, and $Q_n(f)$ are the Fourier transforms of the $n$'th sub-filter, the $n$'th sub-sampled timestream, and the quantization noise in that timestream, respectively. The frequency variable $f$ throughout this section is in units of cycles/sample (or $k/M$, where $M$ is the number of samples being Fourier transformed).

We assume that quantization noise in time samples is uncorrelated, such that $N(f)$ is flat. Moreover, if the PFB channels are narrow enough that the PSD of the underlying signal is approximately flat within one channel, then $\langle |X_n(f)|^2\rangle$ will also be approximately flat \footnote{Sub-sampling the original timestream $x$ by a factor $N$ aliases $N$ Nyquist zones, each $B/n_{chan}$-wide, on top of each other, realizing an approximately flat spectrum.}. Flat PSDs of signal and noise are proportional to their variances, which allows the Wiener filter, $S/(S+N)$, for sub-filter path $n$ to be written exactly as
\begin{align}
\begin{split}
    H_n(f)    &=  \frac{|W_n(f)|^2 \sigma_s^2}{|W_n(f)|^2 \sigma_s^2 + \sigma_n^2}\\
            &= \frac{|W_n(f)|^2}{|W_n(f)|^2 + 1/\mathrm{SNR}}\\
    \label{eq: wiener_derivation}
\end{split}
\end{align}
where $\sigma_n$ is RMS of quantization noise. The dimensionless parameter $\phi^2$ from section~\ref{sec: wiener filter} can now be recognized as $1/\text{SNR}$. For the 4-bit example, the optimal SNR of 19 dB corresponds to $\phi \approx 0.1$. A high value of $\phi$ forces the Wiener filter to use an artificially low SNR, causing signal suppression near channel edges. However, we have found in our numerical experiments that the filtered spectrum is not overly sensitive to the exact choice of $\phi$. Optionally, in \eqref{eq: wiener weights}, we have approximately normalized the filter response by noting that $W_n(f) \sim 1$ for most sub-filters, such that the integral of $H_n(f)$ over $[-0.5, 0.5]$ is roughly $1/(1+\phi^2)$. 

Wiener filtering $\mathbf{d}_n$ followed by decorrelation is equivalent to decorrelation with an effective kernel $H_n(f)/W_n^*(f)$. With the Wiener filter specified by \eqref{eq: wiener_derivation}, the decorrelated estimate (denoted with a hat) of $n$'th timestream's spectrum can be written as
\begin{equation}
    \hat{X}_n(f) = H_n(f) X_n(f) + \frac{H_n(f)}{W_n^*(f)}Q_n(f).
    \label{eq: deconv_spectrum}
\end{equation}
However, of greater practical interest is the spectrum of the reconstituted full-rate timestream, and the bias introduced in its PSD due to quantization noise and Wiener filtering. Conceptually, the full-rate timestream can be obtained by upsampling all $\mathbf{x}_n$'s (appending $N-1$ zeros between their samples), delaying each upsampled $\mathbf{x}_n$ by $n$ samples, and then adding them together. This converts $x_0,0,...$ and $x_1,0,..$ into $x_0, x_1,...$. Following this process, the spectrum of the full-rate timestream can be written in terms of its sub-sampled components as
\begin{equation}
    X(f) = \sum_{n=0}^{N-1} X_n(Nf) e^{-j 2 \pi Nf n}.
    \label{eq: upsampled_spectrum}
\end{equation}
The process of upsampling compresses the original spectrum and creates images, so that full-rate spectrum contains $N$ replicas of the low-rate spectrum in $[-0.5, 0.5]$. A proof of this identity can be found in standard texts on multi-rate signal processing \citep{vaidyanathan2002multirate, rabiner1996multirate}. The relative bias or error in the power spectrum of filtered data is then defined as:
\begin{equation}
\delta(f) = \frac{\langle |X(f) - \hat{X}(f)|^2 \rangle}{\langle |X(f)|^2 \rangle}.
\end{equation}
Using \eqref{eq: wiener_derivation}, \eqref{eq: deconv_spectrum}, and \eqref{eq: upsampled_spectrum}, the bias can be evaluated easily for the case of white input signal. In this case, the spectrum of two sub-filter paths are uncorrelated, such that $\langle X_n(f)X_m(f) \rangle \propto \sigma_s^2 \delta_{mn}$. For the case of naive decorrelation, the relative error is 
\begin{equation}
    \delta(f) = \frac{1}{\mathrm{SNR}}\frac{1}{N} \sum_{n=0}^{N-1} \frac{1}{|W_n(Nf)|^2},
    \label{eq: high-rate bias}
\end{equation}
which shows that the error will increase sharply if one of the sub-filters' response approaches zero. This is illustrated in Fig.~\ref{fig: corrupted_region}. Similarly, relative error in the case of Wiener filtered decorrelation is
\begin{equation}
    \delta(f) = 1 - \frac{1}{N}\sum_{n=0}^{N-1} H_n(Nf).
    \label{eq: high-rate bias}
\end{equation}
Evaluation of $\delta(f)$ in practice involves summation of all $N$ filter responses, followed by $N$ repeats of the summed spectrum, which spans the sampling rate $2B$ of the original timestream. The expression in~\eqref{eq: high-rate bias} is the analytical estimate in Fig.~\ref{fig: wiener comparison}, which lines up well with the Monte-Carlo estimate. It is worth mentioning here that the task of calculating the full-rate spectrum from spectra of $N$ polyphase branches is mathematically identical to calculating the spectrum of interleaved analog-to-digital converters \citep{el2010background}. Finally, near the center of any PFB channel, which corresponds to the region around $H_n(f=0)$ or the bottom edge of Fig.~\ref{fig: eigenvalues}, $H_n \sim 1$, such that \eqref{eq: high-rate bias} can be simplified to
\begin{equation}
    \delta(f)|_{Nf\bmod{N}=0} = \frac{1}{1 + \mathrm{SNR}}.
    \label{eq: high-rate bias simple}
\end{equation}
This simplified expression agrees with the expectation that the relative error in the well-behaved region of is set by maximum achievable quantization SNR, provided the assumptions about signal and noise PSD continue to hold. Since $\mathrm{SNR}\gg 1$, the error in majority of the re-channelized spectra goes as $\sim \mathrm{SNR}^{-1}$, where $\mathrm{SNR}$ increases approximately exponentially with the number of bits. Near the edges of original channels, the exact level of $\mathrm{SNR}$ degradation can be computed from~\eqref{eq: high-rate bias}.

\section{Circulant Inverse PFB Time Complexity}\label{sec: time complexity}

Using the circulant algorithm to invert a chunk of $p$-tap $N$-point PFB spectra of shape (M spectra and N/2 unique channels) requires three stages of FFTs. First stage applies $M$ independent N-point FFTs to invert $\mathcal{F}$, costing $\mathcal{O}(MNlogN)$. This step produces $N$ correlated timestreams $\mathbf{d}_n$'s, each of size $M$. The subsequent decorrelation of these timestreams in Fourier space, according to equation~\eqref{eq: decorrelation}, requires two stages of $N$ independent $M$-point FFTs, with an optional point-wise multiplication step for Wiener filtering, costing roughly $\mathcal{O}(NMlogM)$. Thus, the total computational cost for the circulant inversion scales as $\mathcal{O}(MNlogMN)$. %Since independent FFTs are parallelizable on CPUs or GPUs using existing, well-optimized libraries (like FFTW or CuFFT), and as $M\gg N$ in practice, the overall cost for circulant PFB inversion scales as $\mathcal{O}(MlogM)$. 

The Wiener-filter decorrelation step can also be implemented by passing each of the $\mathbf{d}_n$'s through a finite impulse response (FIR) filter, where the impulse response of the FIR filter is given by the ``decorrelation kernels" shown in Fig.~\ref{fig: convolution kernels}). Cost of FIR filtering one of the sub-sampled timestreams  scales as $\mathcal{O}(ML)$ where $L$ is the length of the decorrelation kernel. The total cost of FIR filtering based IPFB is, therefore, $\mathcal{O}(MN \mathrm{max}(logN, L))$. From Fig.~\ref{fig: convolution kernels}, the length of the longest required FIR kernel (corresponding to sub-filter path $N/2$) is approximately 50 samples for the example case of a 4-tap PFB with 4-bit output quantization (corresponding to $\phi \sim 0.1$). This length increases steadily with higher taps or higher bit-depth. In a typical radio astronomy use-cases, PFB frame sizes, $N$, are in the range $2^{10}-2^{14}$, and the number PFB frames being inverted in a single function call, $M$, might be in the range $10^4-10^6$, such that $log(MN) < L$. Thus, FIR filtering will be slower than the FFT-method by a factor of a few, unless $L$ is artificially reduced by choosing a high value of $\phi$.

Finally, the Wiener-filter machinery can be replicated using a pseudo-inverse solver with Tikhonov regularization for noise suppression. A well-optimized solver can, in principle, solve equation~\eqref{eq: pfbsubsampled} in $\mathcal{O}(M p^2)$ time using banded Cholesky decomposition, for a total cost of $\mathcal{O}(M N p^2)$. With $p=4$ taps, $\mathrm{log_2}M \sim p^2$, such that the circulant algorithm and the Tikhonov solver have similar computational cost. However, the circulant algorithm offers two important benefits. First, its cost does not scale with the number of taps. Due to its $p^2$ scaling, the Tikhonov solver becomes quickly unwieldy for other common configurations like $p=8$ or $16$, and prohibitively expensive for PFB configurations found in telecommunication sector, where the number of taps can be as high as 128. Second, the circulant algorithm allows \textit{any} eigenvalue filtering scheme, not just Wiener filtering. In contrast, custom filtering of singular values of equation~\eqref{eq: pfbsubsampled} requires a full SVD that costs $\mathcal{O}(M^3)$, making it intractable for inverting large amounts of PFB data.

%% For this sample we use BibTeX plus aasjournalv7.bst to generate the
%% the bibliography. The sample7.bib file was populated from ADS. To
%% get the citations to show in the compiled file do the following:
%%
%% pdflatex sample7.tex
%% bibtext sample7
%% pdflatex sample7.tex
%% pdflatex sample7.tex

\bibliography{references}{}
\bibliographystyle{aasjournalv7.1}

%% This command is needed to show the entire author+affiliation list when
%% the collaboration and author truncation commands are used.  It has to
%% go at the end of the manuscript.
%\allauthors

%% Include this line if you are using the \added, \replaced, \deleted
%% commands to see a summary list of all changes at the end of the article.
%\listofchanges

\end{document}